# Non-Hermitian Morphing of Topological Modes

Wei Wang[1], Xulong Wang[1], Guancong Ma[2]

Department of Physics, Hong Kong Baptist University, Kowloon Tong, Hong Kong, China

**Topological modes (TMs) are usually localized at defects or boundaries of a much larger topological lattice[1,2]. Recent studies of non-Hermitian band theories unveiled the non-Hermitian skin effect (NHSE), by which the bulk states collapse to the boundary as skin modes[3–6]. Here, we experimentally demonstrate that the NHSE can conversely affect wavefunctions of TMs by delocalizing them from the boundary. At a critical non-Hermitian parameter, the in-gap TMs even become completely extended in the entire bulk lattice, forming an "extended mode outside of a continuum." These extended modes are still protected by bulk-band topology, making them robust against local disorders. The delocalization effect is experimentally realized in active mechanical lattices in both one-dimensional (1D) and two-dimensional (2D) topological lattices, as well as in a higher-order topological lattice. Furthermore, by the judicious engineering of the non-Hermiticity distribution, the TMs can deform into a diversity of shapes. Our findings not only broaden and deepen the current understanding of the TMs and the NHSE but also open new grounds for topological applications.**

[1]These authors contributed equally to this work.
[2]Email: phgcma@hkbu.edu.hk

Topological matters[1] have generated long-lasting influence across many realms, as evidenced by the vibrant development in topological photonics[7], acoustic and mechanical systems[8]. A hallmark of most topological matters is the existence of TMs localized at the boundaries or interface or defects in a specimen. Imbued with topological protections, many applications such as wave-routing devices[9–11], lasers[12–15], quantum-optics devices[16–18], can benefit from properties like robustness or immunity to backscattering. However, because the existence of the TMs rests on bulk-band topology, any topological application must be built upon a bulk lattice, making them bulky in footprint and costly to fabricate. Here, we demonstrate a non-Hermitian solution to this issue.

Non-Hermitian systems generally have a complex spectrum[19], making them naturally suitable for describing open systems. An open system can exchange energy with the external environment, thereby permitting the presence of effects such as gain and loss, and nonreciprocity, giving rise to a kaleidoscope of novel phenomena[20,21]. In addition, the complex spectrum also entails topology associated with the winding of eigenvalues[22,23] and the exchange of eigenstates around singularities known as exceptional points[24–28]. Research tying non-Hermitian formalism to topological matters led to the discovery of the non-Hermitian skin effect (NHSE)[3–6], by which the bulk states become skin modes localized at open boundaries. NHSE has been experimentally verified in wave systems[29,30], photonics[31], and by quantum walk[32]. Intriguingly, even in the presence of the NHSE, bulk modes restore a Bloch-wave-like extended profile at discrete energies called Bloch points[33,34]. Recently, it was shown that the NHSE can affect the localization properties of the TMs in 1D open lattice[29], and even render them delocalized[35,36], yet experimental confirmation remains absent.

Here, we leverage NHSE to alter the wavefunctions of TMs. The theoretical grounds and proof-of-principal experiments in an active mechanical lattice[29,37] are introduced using a 1D non-Hermitian topological system, in which the complete delocalization of a topological zero mode (TZM) is observed. The effect is extended to 2D topological lattices, wherein we demonstrate the morphing of 1D topological edge mode (TEM) and a 0D second-order topological corner mode (TCM) into diverse profiles by engineering the non-Hermiticity distribution. Although the TMs become extended under the NHSE, they remain spectrally isolated in the bandgap and their topological protections are intact, effectively making them "extended states outside of the continuum" – an opposite of bound states in the continuum[38]. Our findings unshackle TMs from the boundary-localizing characteristics, which can benefit a wide variety of topological applications.

**Delocalization of TZM by NHSE**

We start with a 1D non-Hermitian Su-Schrieffer-Heeger (NH-SSH) chain, as shown in Fig. 1a. The bulk Hamiltonian is

$$H(k_x) = \begin{bmatrix} 0 & v_x + w_x e^{-ik_x} \\ v_x + (w_x + \delta_x)e^{ik_x} & 0 \end{bmatrix}, \quad (1)$$

where $v_x$ and $w_x$ are intracell and intercell hopping coefficients, and $\delta_x$ is nonreciprocal hopping, which is the origin of non-Hermiticity in the system. Here, $v_x, w_x, \delta_x \in \mathbb{R}$ and $v_x = -2.33$, $w_x = -0.38$. The hopping values are retrieved from experimental data. When $\delta_x = 0$, the system is Hermitian and both bulk bands are topologically trivial with a zero topological invariant (quantized Zak phase). When $\delta_x \neq 0$, the system is non-Hermitian, consequently, the energies $E \in \mathbb{C}$ form loops in the complex plane, as shown in Fig. 1b. This indicates the NHSE for the bulk modes under an open boundary condition[39,40]. The NHSE is revealed by examining the wavefunctions of the bulk modes $\psi_n$. In Fig. 1c, we plot $\bar{\psi} = \frac{1}{N}\sum_{n=1}^{N}|\psi_n|^2$, where $N = 60$ is the total number of the states (also the site number). $\bar{\psi}$ collapses towards the left and is vanishing in the bulk. To account for the NHSE, we use the non-Bloch band theory to compute the bulk topological invariants in a generalized Brillouin zone[3]. The topological transition point is at $|v_x| = \sqrt{|w_x(w_x + \delta_x)|}$ (see Supplementary Information section I), so the NH-SSH chain remains topologically trivial in the presence of the NHSE.

We next build an interface with the NH-SSH and a Hermitian SSH (H-SSH) chains, as shown in Fig. 1d. The H-SSH chain has 59 sites and the intracell and intercell hoppings are $w_x, v_x$. It is connected to the NH-SSH chain by an interface hopping $v_d$. It is straightforward to check that the bulk bands of the H-SSH chain have a quantized Zak phase of $\pi$, such that an in-gap TZM is localized at the interface. Chiral symmetry dictates that the TZM remains at zero energy regardless of $\delta_x$ (Fig. 1e). The persistence of the TZM against $\delta_x$ can also be shown by a similarity transformation that turns the system into a Hermitian one (see Supplementary Information section I).

The TZM's wavefunction is dependent of $\delta_x$, as shown in Fig. 1f. When $|\delta_x|$ is small, the TZM remains localized at the interface. As $|\delta_x|$ increases, the TZM gradually delocalizes in the NH-SSH chain. And the TZM becomes an extended mode occupying the entire NH-SSH part at $\delta_x = \delta_{xc} = -1.95$. This extended TZM remains in the middle of the bandgap. Further increase of $|\delta_x|$ results in the TZM to re-localize at the left end of the NH-SSH chain and it dwells in the topologically trivial part of the system.

The NHSE is responsible for such change in the TZM's wavefunction. The nonreciprocal hopping in the NH-SSH chain has the tendency to amplify the local magnitude towards the left side of the lattice, and it effectively counters the exponential decay of the TZM. This can be seen from the characteristic equations in the NH-SSH part,

$$w_x B_{j-1} + v_x B_j = E A_j, \quad v_x A_j + (w_x + \delta_x) A_{j+1} = E B_j, \qquad (2)$$

where $A_j$ and $B_j$ respectively denote the wavefunction components at sublattice A and B of the $j$th unit cell. Since $E = 0$ for the TZM and all $B_j = 0$ because of the bipartite property, the magnitude ratio is

$$\eta = \frac{A_j}{A_{j+1}} = -\frac{w_x + \delta_x}{v_x}. \qquad (3)$$

A decay-free extended TZM appears when $|\eta| = 1$, which corresponds to $\delta_{xc} = v_x - w_x$ (exact non-Bloch $\mathcal{PT}$-symmetry[41]) or $\delta_{xc} = -v_x - w_x$ (broken non-Bloch $\mathcal{PT}$-symmetry). $\delta_{xc}$ also corresponds to the spectral topology transition point for the bulk bands under periodic boundary condition (see Supplementary Information section II). We choose the first $\delta_{xc}$, which guarantees a real spectrum. Equation (3) also shows that the TZM re-localizes at the left end when $|\eta| > 1$, corresponding to $|\delta_x| > |\delta_{xc}|$.

In Fig. 1f, the extended TZM has a spike at the interface, whose height is determined by the characteristic equation across the interface, i.e., $v_x A_I + v_d A_{I+1} = E B_I$, which gives $\epsilon = |A_I/A_{I+1}| = |v_d/v_x|$. This means that the magnitude of the extended TZM is tunable by $v_d$. The spike disappears at $\epsilon = 1$ and the extended TZM becomes entirely flat (Fig. 1g).

**Experimental observation of the delocalization effect**

We experimentally validate the above findings in an active mechanical lattice[29,37]. The building block is a rotational oscillator comprised of a brushless motor and two tensioned springs via a rotational arm (Fig. 2a), which plays the role of the on-site orbital. The oscillators are connected by tensioned springs to form a 1D lattice (Fig. 2b). Nonreciprocal hopping is realized by programmed external actuation of the motors. The systems parameters $v_x$, $w_x$, $\delta_x$, and on-site frequency are retrieved by the Green's function method. (See more details in Methods and Supplementary Information section XII.)

Our experimental lattice consists of a 10-site NH-SSH chain and a 9-site H-SSH chain (Fig. 2b). We excite the lattice at the interface by sending a signal covering 5 – 20 Hz to site-11. We first turn off the nonreciprocity and measure the bulk and interface responses (Fig. 2c). Two peaks separated by a bandgap (the teal curve) are observed in the response of site-8, which corresponds to the two bulk bands and the bandgap. An additional peak is found at 12 Hz when measured at site-9 (the orange curve), which is due to the TZM. We have further measured the rotational response at all sites at 12 Hz. The result is a mode strongly localized at the interface (Fig. 2d, blue). Next, we turn on the nonreciprocal hopping in the NH-SSH chain and measure its effect on the spatial profile of the TZM. In Fig. 2d, we observe that the localization length of the TZM increases with $|\delta_x|$ in the NH-SSH chain. The TZM becomes nearly extended when $\delta_x = -1.88$, which is close to $\delta_{xc}$. The tendency to re-localize is also observed for $\delta_x = -2.45$. By further tuning the spring at the interface to satisfy $\epsilon = 1$, a nearly flat extended TZM is obtained (Fig. 2e). These experimental observations are in excellent agreement with the theoretical results. We also show the results in Supplementary Video 1.

The extended TM can also be realized in a single nontrivial NH-SSH chain, even in the presence of long-range hopping. Extended TM and extended bulk mode can even coexist (see Supplementary Information section II and section III).

Because the NHSE can emerge from a wide range of non-Hermitian systems, our findings offer a recipe for creating TMs beyond the well-known localized, boundary/interface dwelling profile.

Next, we demonstrate such capability by generalizing the phenomenon to two different types of 2D topological systems.

**Morphing of TEM in a 2D stacked topological lattice**

We build a 2D rectangular lattice by stacking the NH-SSH chain along the $y$-direction with an interchain hopping $v_y$, as shown in Fig. 3a. The bulk Hamiltonian reads

$$H_{2D}(k_x, k_y) = \begin{bmatrix} 2v_y \cos(k_y) & V \\ V^* + \delta_x & 2v_y \cos(k_y) \end{bmatrix}, \tag{4}$$

where $V = v_x + w_x e^{-ik_x}$, and $v_x = -0.41$, $w_x = -2.6$, and $v_y = -0.1$. The NH-SSH chain along $x$-direction is nontrivial because of $\sqrt{|v_x(v_x + \delta_x)|} < |w_x|$. The energy spectra of an open lattice having $9 \times 7$ sites are plotted in Fig. 3c, d in which a cluster of TEMs are observed in the bandgap regardless of the values and distributions of $\delta_x$. When $\delta_x = 0$, the TEMs are strongly localized at the left edge, as shown in Fig. 3e. We next turn on the NHSE by setting a nonreciprocal hopping $\delta_x = \delta_{xc} = w_x - v_x = -2.19$ in every unit cell. In Fig. 3f, we see that the TEMs are turned into an extended mode that occupies the entire lattice.

These effects are experimentally observed in the mechanical lattice. The photo in Fig. 3b shows a cropped section of the lattice. The rotational arms are modified into a cross shape for the ease of connecting the coupling springs. We use long-exposure photography to record the real-space profiles of the TEMs. To do so, two phosphorescent stickers are attached to each rotational arm (Fig. 3b). The lattice was placed in darkness and was driven at 12.5 Hz at the left edge. The oscillation traces two arcs under long exposure, where the angles swept by the arcs correspond to the oscillation amplitude. In Fig. 3j, the NHSE is turned off ($\delta_x = 0$) and only the oscillators on the left edge are vibrating, which is the TEMs' response. In Fig. 3k, we set $\delta_x = -2.14$ and observe that every unit cell is oscillating with a near-identical amplitude, indicating the delocalization of the TEMs. The measured profiles conform excellently with the tight-binding results (Fig. 3e, f).

By further engineering the distribution $\delta_x$ as functions of $y$, a wide variety of profiles for the TEMs can be obtained. Three examples are presented in Fig. 3g-i. Here, the NHSE has different strengths in each NH-SSH chain, by which the response field is morphed into a pyramidal shape (Fig. 3g, l), a triangle (Fig. 3h, m), and a V shape (Fig. 3i, n). These effects are further shown in Supplementary Video 2. Herein, $\delta_x$ for each chain is determined by considering the localization length. We first obtain the Hermitian TZM's localization length $\xi_H = (\ln|w_x/v_x|)^{-1}$. The targeted localization length in the presence of $\delta_x$ is

$$\xi_{NH} = [\ln|w_x/(v_x + \delta_x)|]^{-1}, \tag{5}$$

by which we can obtain $\delta_x = w_x(e^{-1/\xi_{NH}} - e^{-1/\xi_H})$. The NHSE can also be produced in the $y$-direction, by which the TEM can be further collapsed to a corner (Supplementary Fig. 10).

**Morphing of TCM by higher-order NHSE (HO-NHSE)**

The HO-NHSE is a recently discovered non-Hermitian phenomenon by which bulk states in a lattice collapse to 0D corner skin modes. Here, we conversely leverage the HO-NHSE to manipulate the TCM in the 2D topological quadrupole insulator. Figure 4a shows the unit cell, and the bulk Hamiltonian reads

$$H_Q(k_x, k_y) = \begin{bmatrix} 0 & R_x(k_x) & R_y(k_y) & 0 \\ M_x(k_x) & 0 & 0 & -R_y(k_y) \\ M_y(k_y) & 0 & 0 & R_x(k_x) \\ 0 & -M_y(k_y) & M_x(k_x) & 0 \end{bmatrix}, \tag{6}$$

where $R_\mu(k_\mu) = v_\mu + w_\mu e^{-ik_\mu}$ and $M_\mu(k_\mu) = (v_\mu + \delta_\mu) + w_\mu e^{ik_\mu}$ with $\mu = x, y$, $v_y = -0.49$, $w_y = -2.12$. Note that to generate HO-NHSE, nonreciprocal hopping terms are required for both $x$ and $y$ directions. Here, the hopping terms, including the nonreciprocal ones, are arranged to preserve the $\pi$-flux within each four-site plaquette, thereby giving a quantized non-Hermitian generalization of quadrupole moment (see Supplementary Information section VII). Also, despite the presence of non-Hermiticity, equation (6) respects chiral symmetry: $\Gamma H_Q \Gamma^{-1} = -H_Q$ with $\Gamma = \text{diag}(1, -1, -1, 1)$. Hence the spectrum is always symmetric about zero energy – a characteristic that is carried over to the open lattice, as shown in Fig. 4d. Here, because the open lattice has an odd number of sites, the system only has one TCM pinned at zero energy, appearing at the lower-left corner, as shown in Fig. 4e. Next, we engage the nonreciprocal hopping and examine how they affect the TCM. First, we only turn on $\delta_x$ and set it to $\delta_x = \delta_{xc} = w_x - v_x = -2.19$ and keep $\delta_y = 0$, so that the NHSE only exists along the $x$ direction. As a result, the TCM is delocalized along the lower edge of the lattice, effectively becoming an edge state. The wavefunction is shown in Fig. 4f. Similarly, the TCM can also delocalize along the left edge when $\delta_y = \delta_{yc} = w_y - v_y = -1.63$ and $\delta_x = 0$ (not shown here). When nonreciprocal hoppings along both directions are nonzero and $\delta_x = \delta_{xc}, \delta_y = \delta_{yc}$, the HO-NHSE takes place and morphs the TCM into an extended state occupying the entire lattice (Fig. 4g).

These effects are also experimentally realized. The unit cell is shown in Fig. 4b. Here, positive hopping terms are implemented by crossing the springs. The vibrational profiles of the TCM under different non-Hermitian conditions are shown in Fig. 4h-j. Excellent agreement with theoretical predictions is obtained. These results are also shown in Supplementary Video 3.

**Discussion**

There is an alternative understanding of the NHSE's influences on the TMs when the nonreciprocal hopping is uniform: the 2D Hamiltonians [equations (4) and (6)] can be decomposed into 1D models. We also present a theoretical study on a Kagome lattice in Supplementary Information section X, which cannot be decomposed into lower-dimensional models. For these models, chiral symmetry or its generalized version is preserved, which ensures the TMs pinned to zero energy. However, this is not a necessary condition. The delocalization can be achieved without chiral symmetry (see Supplementary Information section II).

The NHSE is known to collapse the bulk states towards open boundaries. Our work achieves the opposite by converting localized TMs into extended modes of unconventional shapes while preserving the topological characteristics (the robustness of the extended TMs is discussed in Supplementary Information section IV and section V). The morphing effect is reconfigurable by reprograming the nonreciprocity. The extended TMs have no Hermitian counterpart and the underlying physics is general. Our findings show that non-Hermiticity is a powerful new degree of freedom by which the wavefunction of TMs can be tailored to suit different needs. Because both topological matters and nonreciprocity have been extensively studied in photonics[7,42] and acoustics[8,43], the NHSE's influences on TMs should be realizable in these systems. We expect the extended TMs to bring unprecedented opportunities for topological manipulations of waves and light, and can be the foundation for, e.g., a large-area single-mode topological laser and a coherent topological beam splitter (see Methods).

## Methods

### The mechanical lattice

Our experimental systems are lattices of coupled active rotational harmonic oscillators. The elementary building block is comprised of a brushless DC motor (LDPOWER 2804) with a rotational arm attached to it (Supplementary Fig. 14). In the 1D experiments, the arms are made of plastics and have a length of 129.5 mm. In 2D lattices, the cross-shaped arms are made of aluminum alloy and have a length of 129.5 mm and a width of 93.6 mm. In both cases, the arm has two large holes drilled on the ends, through which weights (e.g., bolts and nuts) can be inserted to tune the moment of inertia. The arm is attached with two identical anchoring springs with a spring constant of 31.35 N/m, which provides restoring torques. The resonant frequency of an isolated oscillator is thus determined by the spring constant and the moment of inertia. To form a lattice, multiple oscillators are coupled by additional tensioned springs connecting the arms of the neighboring units. The hopping magnitude is tunable either by choosing different spring constants or by adjusting the connecting position on the arms (Supplementary Fig. 15). Naturally, stronger spring and (or) longer force-arm results in larger hopping magnitude. The sign of the hopping coefficients can also be chosen. Our analysis shows that the hopping sign is negative (positive) when two neighboring arms are connected by springs that are parallel (crossed). See Supplementary Information section XI for details.

The instantaneous angular displacement $\theta(t)$ of each oscillator is measured by a magnetic rotary position sensor (AMS AS5047P). To do so, we firmly attach a diametrically magnetized magnet to the shaft of the motor. The sensor is placed 2 mm below the magnet. The measured signals are digitalized by a microcontroller (Espressif ESP32) with a sample rate of 1 kHz and then transferred to a computer. The received signals are not only the measurement of local response but also used as feedback to generate nonreciprocal hopping.

### Realization of nonreciprocal hopping

The motor can be actuated by a driver chip (STMicroelectronics L6234PD), which is programmed to generate a response torque based on the measured $\theta(t)$ from the neighboring site. Take the nonreciprocal hopping $\delta_x$ between site $B_{I-1}$ and site $A_I$ in Fig. 1d as an example (Supplementary

Fig. 16). $\theta_I^A(t)$ is monitored in real time and the motor of site $B_{I-1}$ is instructed to generate a torque $\tau_{I-1}^B(t) = \alpha\theta_I^A(t)$, where $\alpha$ is a constant that tunes the nonreciprocity strength. Meanwhile, no active torque is applied to the motor of site $A_I$. Together with the coupling spring between the two sites that produce $w_x$, the net effect is the hopping from $A_I$ to $B_{I-1}$ being $w_x + \delta_x$, while the hopping from $B_{I-1}$ to $A_I$ being $w_x$. The implementation of the nonreciprocal intracell hopping in 2D topological lattices is achieved in a similar manner (Supplementary Figs. 17, 18).

**Acknowledgments.** The authors thank C. T. Chan, Zhao-Qing Zhang, Kun Ding, Ruo-Yang Zhang for discussions, and Qiyuan Wang for assisting with the experiments. This work was supported by the National Natural Science Foundation of China (11922416), the Hong Kong Research Grants Council (12302420, 12300419).

**Author contributions.** W.W. developed the theory and performed numerical calculations. W.X. developed the experimental platform. W.W. and G.M. designed the experimental systems. W.W. and W.X. carried out the measurements. All authors analyzed the results. W.W. and G.M. wrote the manuscript. G.M. led the research.

**Competing interests.** The authors declare no competing interests.

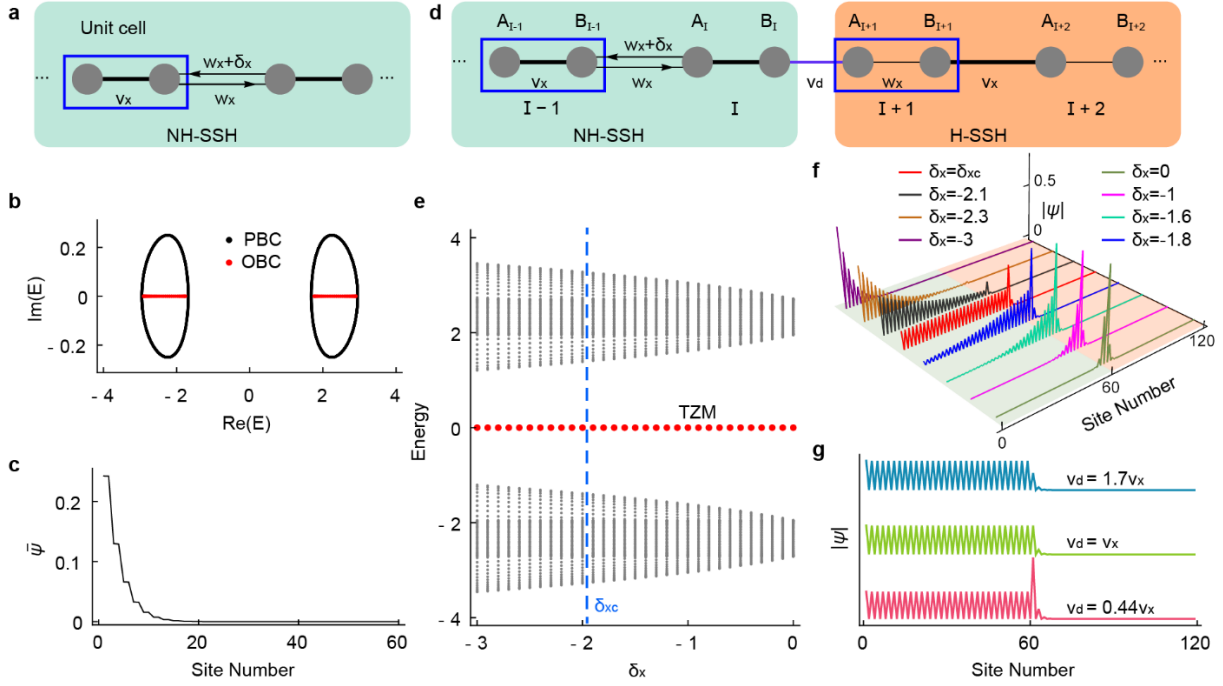

**Fig. 1 | Delocalization of TZM by NHSE. a**, A 1D NH-SSH chain with nonreciprocal intercell hopping $w_x$ and $w_x + \delta_x$. The blue box encloses the unit cell. **b**, Complex energy of the NH-SSH chain under periodic boundary condition and open boundary condition. The chain has 60 sites with $v_x = -2.33$, $w_x = -0.38$, $\delta_x = -0.5$. **c**, Spatial distribution $\bar{\psi}$ of all the wavefunctions in an open chain shows the NHSE. **d**, An interface formed by a topologically trivial NH-SSH chain and a topological H-SSH chain. **e**, The energy spectra of the interface system shown in **d** and their dependence on $\delta_x$. Here, $v_d = -1.03$. The red dots mark the TZM which is pinned at zero energy for all values of $\delta_x$. **f**, The real-space wavefunctions of the TZM for different $\delta_x$. It is seen that the increase of $|\delta_x|$ delocalizes the TZM until $\delta_x = \delta_{xc} = -1.95$, at which the TZM becomes fully extended in the NH-SSH chain. The TZM re-localizes at the left end of the system for even larger $|\delta_x|$. **g**, When $\delta_x = \delta_{xc}$, tuning $v_d$ can produce a constant-amplitude mode in the NH-SSH chain.

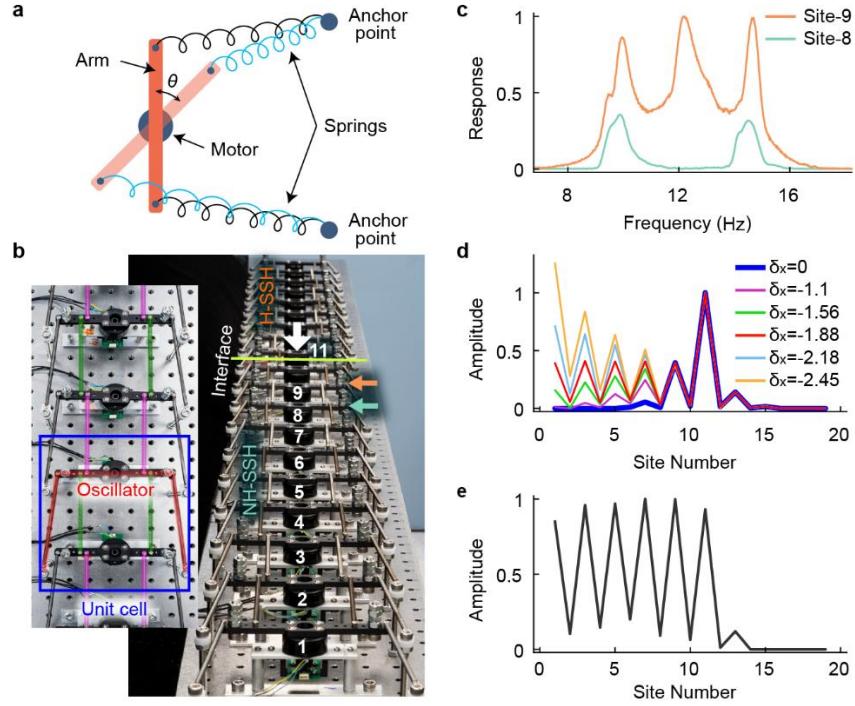

**Fig. 2 | Experimental observation of the delocalization effect. a**, The schematic drawing of an isolated rotational oscillator, which serves as on-site orbital in our experimental lattices. **b**, A photo of the 1D NH-SSH and H-SSH interface system. The white numbers label the sites. The TZM is excited at site-11, which is marked by the white arrow. The green and orange arrows are the measurement positions of the data shown in **c**. The left inset is a top-down view of the lattice. The coupling springs are colored for clarity. The blue box encloses one unit cell. **c**, Response spectra measured at site-8 and site-9 with excitation at site-11. The TZM response is seen in the orange curve (site-9). **d**, Measured oscillation amplitudes of the TZM for different values of $\delta_x$. The TZM is nearly extended at $\delta_x = -1.88$. The excitation is at site-11 at 12 Hz. **e**, Measured oscillation amplitude of the near-constant-amplitude TZM ($v_d = v_x$).

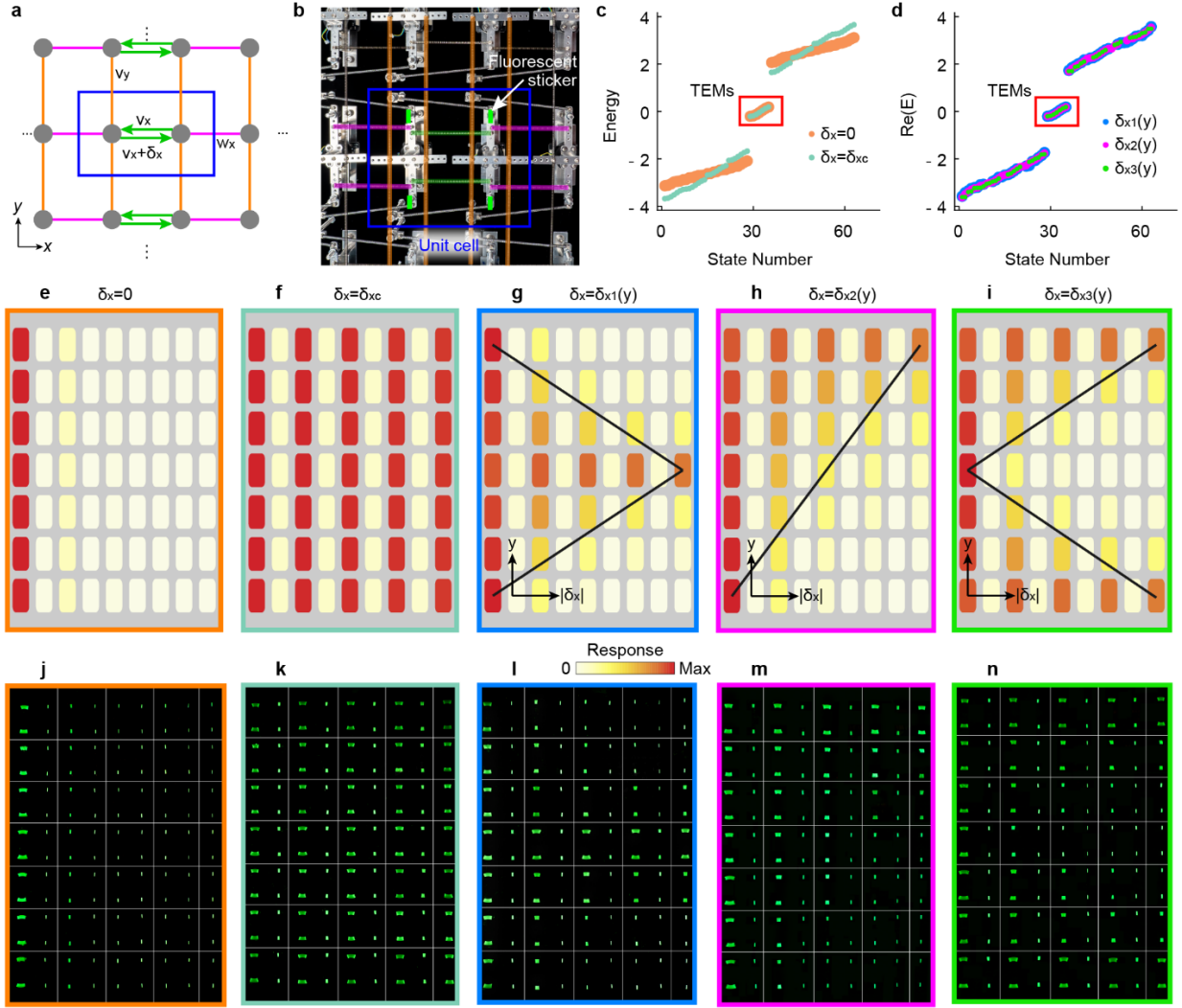

**Fig. 3 | Morphing of TEM in a 2D stacked topological lattice. a**, The schematic drawing of the 2D stacked topological lattice. The blue box marks a unit cell. **b**, A photo of the experimental lattice realizing **a**. The coupling springs are colored for clarity. **c**, The energy spectra of a $9 \times 7$-site lattice with different nonreciprocal hopping in the $x$-directions. Here, $v_x = -0.41$, $w_x = -2.6$, and $v_y = -0.1$. The TEMs are marked by the red box. **d**, The real part of the energy spectra of the open lattice with three different nonreciprocal hopping distributions. (Note that the energies are complex here.) **e-i**, The theoretical response fields with lattice excited at the left edge are plotted in $\delta_x = 0$ (**e**), $\delta_x = \delta_{xc}$ (**f**), and nonuniform distribution of $\delta_x(y)$ (**g-i**) as indicated by the overlaying curves. **j-n**, The long-exposure photos showing the oscillation profiles in the lattice with harmonic excitation at 12.5 Hz at the left edge. The localized Hermitian TEM (**j**) becomes an

extended mode over the entire lattice in **k**. It also deforms to pyramidal **(l)**, triangular **(m)**, and V shape **(n)** by different $\delta_x(y)$. The grids in **j-n** mark the unit cells.

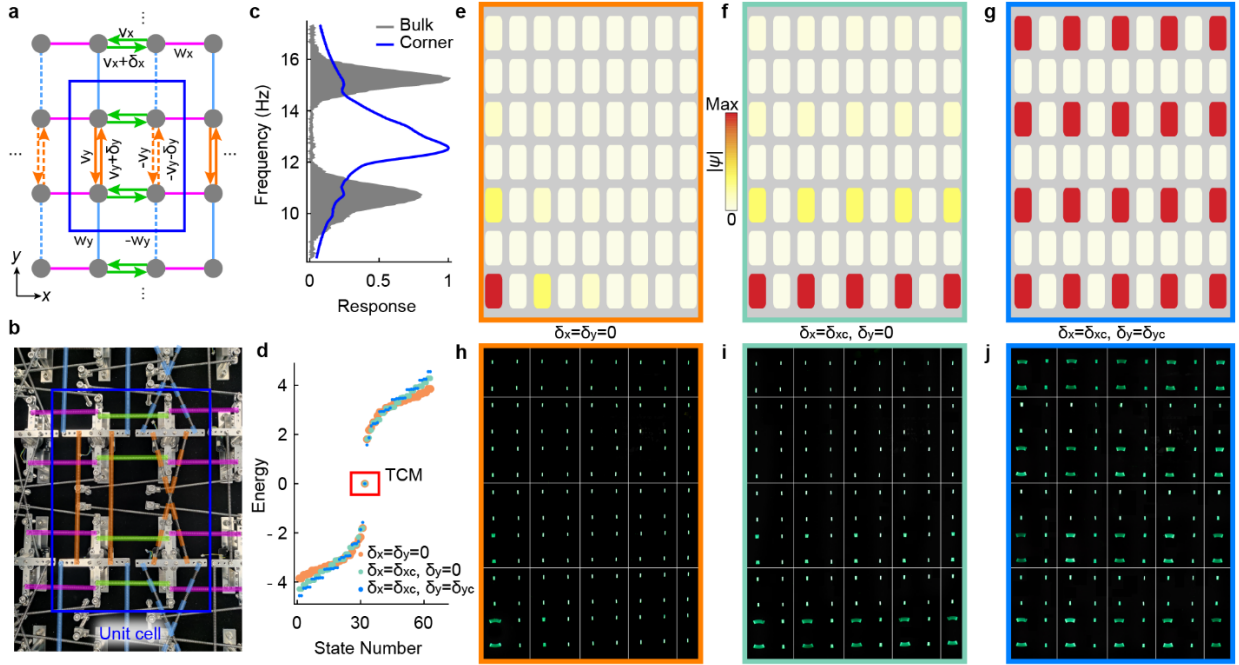

**Fig. 4 | Morphing of TCM by HO-NHSE. a**, The model of a non-Hermitian topological quadrupole insulator. **b**, A photo of the experimental lattice realizing **a**. **c**, The measured response spectra of the TCM at the lower-left corner of the Hermitian topological quadrupole insulator. The gray shaded region corresponds to the response of the bulk modes. **d**, The energy spectra of an open lattice consisting of $9 \times 7$ sites with $\delta_x = \delta_y = 0$ (orange), $\delta_x = \delta_{xc}$, $\delta_y = 0$ (teal), and $\delta_x = \delta_{xc}$, $\delta_y = \delta_{yc}$ (blue). **e-g**, The wavefunctions of the TCM when $\delta_x = \delta_y = 0$ (**e**), $\delta_x = \delta_{xc}$, $\delta_y = 0$ (**f**), and $\delta_x = \delta_{xc}$, $\delta_y = \delta_{yc}$ (**g**). **h-j**, The measured vibration fields corresponding to the TCM (**h**), the extension of the TCM along the lower edge (**i**), and into the 2D surface (**j**) under the harmonic excitation at 12.4 Hz at the corner. In **i, j**, $\delta_x = -2.18$ and $\delta_y = -1.7$. The grids in **h-j** mark the unit cells.

18